**Hypothesis Article**

# Eta-Earth Revisited I: A Formula for Estimating the Maximum Number of Earth-like Habitats

Helmut Lammer[1*], Manuel Scherf[1,2], Laurenz Sproß[1,2]

[1] Austrian Academy of Sciences, Space Research Institute, Graz, Austria
[2] Institute of Physics, University of Graz, Austria

*Email Corresponding Author: helmut.lammer@oeaw.ac.at

**Abstract**

In this hypothesis article, we discuss the basic requirements of planetary environments where aerobe organisms can grow and survive, including atmospheric limitations of millimeter-to-meter-sized biological animal life based on physical limits, and $O_2$, $N_2$, and $CO_2$ toxicity levels. By assuming that animal-like extraterrestrial organisms adhere to similar limits, we define Earth-like Habitats ($\eta_{EH}$) as rocky exoplanets in the Habitable Zone of Complex Life that host $N_2$-$O_2$-dominated atmospheres with minor amounts of $CO_2$, at which advanced animal-like life can in principle evolve and exist. We then derive a new formula that can be used to estimate the maximum occurrence rate of such Earth-like Habitats in the Galaxy. This contains realistic probabilistic arguments that can be fine-tuned and constrained by atmospheric characterization with future space and ground-based telescopes. As an example, we briefly discuss two specific requirements feeding into our new formula that, although not quantifiable at present, will become scientifically quantifiable in the upcoming decades due to future observations of exoplanets and their atmospheres.

## 1. Introduction

Since the discoveries of low mass exoplanets (i.e., sub-Neptunes, super-Earths, etc.) due to the CoRoT (e.g., Deleuil et al., 2011) and Kepler (e.g., Borucki et al., 2011) space observatories, an understanding of the frequency of potentially habitable Earth-like exoplanets is a major area of interest (e.g., Catanzarite and Shao, 2011; Dressing and Charbonneau, 2013; Gaidos, 2013; Petigura et al., 2013; Batalha, 2014; Silburt et al., 2015; Barbato et al., 2018; Zink and Hansen, 2019; Westby and Conselice, 2020; Bryson et al., 2021). For finding Earth-size and/or Earth-mass exoplanets that orbit inside their host star's habitable zone (HZ), which allows the existence of liquid water (e.g., Hart, 1979; Lammer et al., 2009; Kasting et al., 1993; Kopparapu et al., 2010), more space observatories such as TESS (Ricker et al., 2014), CHEOPS, JWST, PLATO (Rauer et al., 2014), ARIEL (Tinetti et al., 2018), the proposed observatories LIFE (Quanz et al., 2022), and HWO[1] and instruments on the Extremely Large Telescope (ELT) will be in place to determine the physical and orbital characteristics of promising planets, thereby focusing on the composition of a possible atmosphere.

---

[1] See, e.g., https://cor.gsfc.nasa.gov/studies/habitable-worlds/hwo.php (website visited on June 27, 2023).

Whether aerobic organisms that can evolve in an oxygenated environment, including complex animal-like life as we know it from Earth, also exists outside the Solar System is closely related to the existence of Earth-like habitats. This opens up two fundamental questions (e.g., Fridlund, 2000):

- How common is Earth as a planet?

- How common is life in the Galaxy/Universe?

To date, Earth is the only known example of a planetary habitat with a biosphere where life itself originated and complex aerobic animal life evolved. The potential number of inhabited exoplanets is linked to the term Eta-Earth (also written as $\eta_\oplus$), which is only loosely defined as the mean number per star of rocky exoplanets (with a study-dependent radius ranging mostly somewhere between $0.5R_{pl} \leq R_\oplus \leq 2R_{pl}$) that reside in the optimistic habitable zone (HZ) of their host star (e.g., Haghighipour, 2014; Traub, 2015). Astronomers who apply the radial velocity method for detecting masses of exoplanets sometimes also relate $\eta_\oplus$ to the occurrence rate of planets within a mass range of $M_{pl} < 10M_\oplus$ (e.g., Wittenmyer et al., 2011). The variable $\eta_\oplus$ is similar to $n_e$, which is one of the key factors that enters the Drake equation (Sagan, 1963; Vakoch and Dowd, 2015)

$$N_{civ} = R \times f_{pl} \times n_e \times f_{life} \times f_{in} \times f_{civ} \times L, \qquad (1)$$

that gives $N_{civ}$, the number of hypothetical extraterrestrial civilizations in the Galaxy that communicate through electromagnetic waves. The equation traces back to astronomer and astrophysicist Frank Drake who searched for a way to stimulate scientific dialogue at the first meeting of the Search for Extraterrestrial Intelligence (SETI) in 1961 (Carigi, 2015; Sagan, 1963). The value $R$ corresponds to the average rate of star formation in the galaxy, $f_{pl}$ is the fraction of stars with planets, $n_e$ is the number of habitable planets per planet-hosting star, $f_{life}$ is the fraction of those planets that host life, $f_{in}$ is the fraction of such planets with intelligent life, and $f_{civ}$ is the fraction of intelligent life that releases detectable signs of their existence into space (i.e., civilizations). Finally, $L$ is the timespan over which such civilizations release these signals into space. The main point of criticism of the Drake Equation is that estimated values for several of its factors remain highly conjectural and arbitrary.

Though there may be many terrestrial planets in orbits within the HZ of their host stars that are potentially habitable, by definition, such as those discovered in the TRAPPIST-1 system (e.g., Gillon et al., 2017) that still does not necessarily mean that they are inhabited. Therefore, a crucial question concerning $\eta_\oplus$, in addition to "Are they habitable?" is "Could life ever arise there and, if so, could it evolve toward aerobe lifeforms, including complex animal life or even an intelligent technological civilization?" Some major problems with planets being covered by the common definitions of $\eta_\oplus$ are their unknown accretion histories, their internal composition and volatile content, the importance of continents, oceans, and tectonic regimes (e.g., Stern and Gerya, 2024), and in addition to that connected carbon-silicate and nitrogen cycles, and other relevant parameters such as the planets' distributions of radioactive heat-producing isotopes (Birmingham et al., 2020; Lammer et al., 2020a; 2021; O`Neill et al., 2016; 2020; Wang et al., 2020; Erkaev et al., 2023).

Furthermore, the billion-year evolution of planetary atmospheres of discovered $\eta_\oplus$-planets, including the evolving radiation and plasma environment of their host stars (Guinan and Ribas, 2002; Ribas et al., 2005; Johnstone et al., 2021, 2015; Tu et al., 2015), will also remain unknown to its discoverers but have a huge influence on the possible arising of lifeforms. These parameters are not directly considered in the Drake equation and in the common definitions of $\eta_\oplus$. Another problem with the Drake Equation follows from the subsequent variables $f_{life,}$ $f_{in}$, $f_{civ,}$ and $L$ for which no reliable quantitative estimates are available at present. In other words, any result obtained by the Drake formula is based on pure speculation and very easily biased unintentionally by its users. In the following, we propose a new approach that **focuses** on so-called Class I habitats[2] (Lammer et al., 2009; Lammer 2013) based on:

i) scientifically quantifiable parameters important for the evolution and long-term stability of Earth-like planets with an $N_2$-$O_2$-dominated atmosphere and,

ii) the characterization/detection of their planetary atmospheric bulk gases.

This allows a renewed definition of the crucial parameter $\eta_\oplus$ while not strictly **focusing on** the size or mass of a terrestrial exoplanet within the HZ of its host star **or** on various biosignature trace gases that have been proposed and studied over the last decades (e.g., Schwieterman et al., 2018; and references therein) such as $CH_4$ (e.g., Sagan et al., 1993; Krasnopolsky et al., 2004; Rugheimer et al., 2013; Airapetian et al., 2017), $CH_3Cl$ (Segura et al., 2005; Rugheimer et al., 2015), $NH_3$ (Seager et al., 2013a, 2013b), sulfur gases and $C_2H_6$ (Pilcher, 2003; Domagal-Goldman et al., 2011), and organic hazes (Arney et al., 2016, 2017).

Here, we proceed from three essential atmospheric species, namely $N_2$, $O_2$, and $CO_2$, each within a certain range of partial surface pressures, for convenience called "air" in the following, and their importance for aerobic complex lifeforms. Aerobic organisms can survive and grow in an oxygenated environment. On Earth, $O_2$ built up in the atmosphere during a period that is called the Great Oxidation Event (GOE) $\approx 2.4 - 2.1$ Gyr ago (Lyons et al., 2014), which transformed the existing lifeforms and environments afterward (Olejarz et al., 2021).

---

[2] Class I habitats represent rocky planets where stellar and geophysical conditions allow Earth-like planets with $N_2$-$O_2$-dominated atmospheres (Earth-like Habitats: EHs) to evolve, so that **aerobic organisms',** including complex multicellular animal-like life forms may originate and inhabit the planet's hydrosphere, surface and subsurface environments.

In the definition of aerobic complex lifeforms, we follow the works of Catling et al. (2005), Schwieterman et al. (2019), and Ramirez (2020) and rather than focus on aerobe prokaryotic bacteria and archaea, we focus on "advanced metazoans," a specific niche of eukaryotes that can be defined as millimeter- to meter-sized tissue-grade animals with blood vascular systems; or in simple language, we focus on "complex animal life" (CAL). One should keep in mind, however, that other forms of complex life such as fungi and plants, which are not subsumed within this definition, may also be limited by the chemical consequences of the conventional habitable zone (Schwieterman et al. 2019).

We also point out that the outcome of the terrestrial evolutionary pathway that led to CALs on Earth may not necessarily be inclusive of other evolutionary pathways that may exist on EHs in the Galaxy. Such pathways may, for example, provide unknown compensatory mechanisms for lifeforms that live under extreme environmental conditions that are dangerous or fatal for CALs on Earth. Any alternative evolutionary scenarios might, therefore, offer fewer environmental constraints than CALs as we know it. Moreover, we also point out that our hypothesis is only valid for biological lifeforms but should not be applied to "hypothetical" ETI-lifeforms that have been overtaken or transcended by inorganic intelligence like artificial intelligent machines as recently discussed in the work of Rees and Livio (2023). It should also not be applied to ETIs that spread from an Earth-like Habitat to planetary bodies other than EHs. Furthermore, we do not speculate on hypothetical complex lifeforms that may evolve on other potential habitats such as subsurface oceans of icy satellites and Titan-like bodies outside the classical HZ (Class III and IV habitats, Lammer et al. 2009). Such an alternative biochemistry is scientifically viable to investigate though at the present time, not proven to exist (Davila and McKay, 2014).

In Section 2, we discuss the connection between $N_2$-$O_2$-dominated atmospheres and biospheres with complex aerobic lifeforms, while we derive a formula for Earth-like Habitats in Section 3. Section 4 concludes the article. The second article of our study, Scherf *et al.* (2024, this issue), then applies our novel formula to derive such maximum number of EHs in the galactic disk given the present stage of knowledge.

## 2. The link between $N_2$-$O_2$-dominated atmospheres and biospheres

The classical concept of the circumstellar habitable zone was first introduced by Huang (1959) and has been further developed by several authors (e.g., Rasool and deBergh, 1970; Hart, 1979; Kasting et al., 1993; Kopparapu et al., 2013; 2014; Kopparapu et al., 2016). Within the HZ of a main sequence star, rocky planets can host liquid water and stabilize their climate systems through the carbon-silicate cycle.

In the classical HZ, the inner edge is determined by the stellar luminosity that prevents a runaway or moist greenhouse effect, while the outer edge of the HZ is usually constrained by an upper $CO_2$ limit where higher $CO_2$ levels can compensate for the decreasing stellar luminosity (e.g., Kasting et al., 1983; Kopparapu et al., 2013). The evolvement of such conditions is related to a complex interplay of $H_2O$ and $CO_2$ with a long-time active carbon-silicate cycle and leads to an estimated HZ range of ~0.95 – 1.7 AU (e.g., Kopparapu et al. 2013). By examining the toxicity limits of $CO_2$ partial pressures for complex, animal-like lifeforms, Schwieterman et al. (2019) recently defined an even smaller range for an HZ for Complex Life (HZCL), which is about 0.95 – 1.2 AU for solar-like stars. Following this study, Ramirez (2020) applied to the HZCL-concept of Schwieterman et al. (2019) an energy balance model based on constraints of $CO_2$ and, for the first time, atmospheric $N_2$ related to the respiration limits of CALs, by assuming that advanced lifeforms experience comparable physiological limitations such as air-breathing on EHs. According to Ramirez (2020), the theoretical $N_2$-respiration limit for adult humans is at $\approx 2$ bar, while newborns have a lower $N_2$ tolerance level of $\approx 1.3$ bar. Eventually, these authors obtained a range of $\approx 0.95 - 1.31$ AU for the HZCL of solar-like stars. Although CALs might theoretically evolve to tolerate different toxicity limits on EHs other than Earth, we follow the suggestions of Schwieterman et al. (2019) and Ramirez (2020) as good guide values for atmospheric environments. From these two very recent studies, one can conclude that large fractions of atmospheres within a classical HZ are hostile to aerobic complex animal life.

Based on several studies designed to investigate the interconnection between microbial life and atmospheric $N_2$ (Stüeken et al., 2016a; 2016b; 2020; Lammer et al., 2019; Sproß et al., 2021), one can conclude that the detection of an exoplanet with an $N_2$-$O_2$-dominated atmosphere and a minor amount of $CO_2$ orbiting within the HZCL will be the best indication for inhabitation by microbial lifeforms that are able to perform processes similar to denitrification and anammox. From our definition of CALs in Sect. 1 and the findings of the above-mentioned studies by Schwieterman et al. (2019) and Ramirez (2020) together with further studies that focus on the relevance of $O_2$ (e.g., Catling et al. 2005a; 2005b), one can further conclude that CALs can eventually evolve on such planets. Any additional detection of gases like $H_2O$, $CH_4$, and other biologically relevant gases would support this indication.

### 2.1 Basic requirements and limits for large aerobic complex lifeforms

If one compares the abundance of elements within the measurements of elemental distributions from stars, interstellar clouds, comets, meteorites, our Sun, and the known planets in the Solar System, one finds the following 10 elements to be the most abundant: $H_2$, He, $O_2$, C, Ne, N, Mg, Si, Fe, S (e.g., Smith, 1982); even though, the depletion of some elements on particular Solar System bodies must be mentioned (e.g., Wang et al. 2019; Benedikt et al. 2020; Lammer et al. 2020b and references therein). The most common elements in the universe are $H_2$ and He; they comprise $\approx 98\%$ of all elements with 75% $H_2$ and $\approx 23\%$ He. The next elements in the sequence, O, C, and N, are up to $\approx 10^3 - 10^4$ times less abundant but are the fundamental building blocks of life, as we know it. All other elements are much rarer than these, which shows that life on Earth is based on the most abundant, rather than on some low abundance, elements in the Galaxy. The acronym CHON corresponds to the four most common elements in living organisms: C, H, O, and N. The six most biological molecules in aerobic lifeforms on Earth are often referred to as CHNOPS. Phosphorus (P) containing phosphates are an essential component of RNA and DNA and are required to form main molecules used as energy powering the cell in all lifeforms (Campbell et al., 2006). Sulfur (S) is contained in the amino acids' cysteine and methionine and therefore necessary for protein synthesis and structure (e.g., Brosnan and Brosnan, 2006); it also plays a crucial role in sulfur-containing antioxidants and aids in reducing DNA, RNA, and protein damage in oxidative environments (e.g., Mukwevho et al. 2014). In the Solar System, chondrites are the most common types of meteorites and about 5% of them belong to carbonaceous chondrites, which are rich in CHNOPS elements and frequently collided with Earth during its early history (Raymond and Morbidelli, 2000).

The availability of CHNOPS elements in the Galaxy is the first and most obvious reason why complex aerobic extraterrestrial life should also be based on these elements. However, CALs can only evolve and exist in an appropriate habitat and are therefore dependent on the number of $\eta_{EH}$. One can expect that the evolutionary pathway that life on Earth trod is most likely not the only possibility for life. However, because of the universal abundance of the available building blocks of life, but also due to our limited knowledge about any other potential kinds of life, it makes sense to consider similar forms of biological complexity as known from our singular sample, that is, CALs interlinked to an $N_2$-$O_2$-based atmosphere with minor amounts of $CO_2$ in combination with liquid $H_2O$. Terrestrial planets within the classical water-based HZ are regarded as the main targets for future biosignature searches because the surface temperatures allow a significant exchange of gases between the surface and atmosphere (e.g., Kasting, 2014; Schwieterman et al., 2018; 2019). In the following, we will briefly discuss the role of water, oxygen, $CO_2$, and nitrogen in more detail.

### 2.1.1 Water

The first basic requirement for the origin and evolution of aerobe lifeforms, including millimeter-to-centimeter-size CALs, is liquid water on a terrestrial exoplanet due to its exceptional capability as a solvent (e.g., Brack 2002; Schulze-Makuch and Irwin, 2004; Lammer et al., 2009; Westall and Brack, and references therein). Water's versatility and adaptability support important chemical reactions. Its molecular structure is related to the maintenance of important shapes for cells' inner components and their outer membranes. Because of these important capabilities, $H_2O$ makes up the highest percentage of the body weight of large animals, including humans with up to 60% (Sheng and Huggins, 1979). Within large ranges of temperature and pressure, water stays liquid and can provide the following main characteristics that are known to be indispensable for lifeforms to evolve:

***A large dipole moment that allows strong hydrogen bonds:*** When one atom is more electronegative than another atom and the difference of the electronegativity vector points in the same direction, the atom pulls more tightly on the shared pair of electrons as a result. In water molecules, every O atom is slightly negatively changed about the H atoms. Through its angular structure, the charge difference creates a polarity that allows strong bonds between $H_2O$ molecules or to any other polar molecules such as CO or $NH_3$ (see e.g., Maréchal 2007). This characteristic enables water to be a strong polar organic solvent, which allows the metabolism of life as we know it.

**To stabilize macromolecules**: Water molecules have the important property to contribute to biomolecule formation and their stability, dynamics, and functions (Alberts et al., 2020; Mitusińska et al., 2020). The $H_2O$ molecules are tiny enough to penetrate the core of macromolecules such that their native structure can be stabilized or destabilized (Okumura et al., 2021). For example, it can also participate in processes that occur in a protein's core (e.g., Papoian et al., 2004).

**To orient hydrophobic-hydrophilic molecules**: Biomolecules form molecules from H-bonds with water molecules and are important to the adjustment of properties, such as surface charge, structural flexibility, and hydrophilicity. For example, lipid bilayer membranes are based on hydrophobic interactions between amphiphilic molecules that constitute an interface where water is attracted to the outsides of the membrane (hydrophilic) but is prevented from going through the interior layer (hydrophobic) in aqueous systems (Watanabe et al., 2020), which likely has been very important for forming the first protocells (e.g., Deamer, 2017).

**The negative thermal expansion of water:** Since $H_2O$ has its maximum density at 277,15 K (4˚ C), frozen water will stay at the surface, and a stable habitable environment can be maintained at the bottom. If this anomaly did not exist, circulation would lead to the freezing of the entire liquid. Another problem would be that freezing at the bottom would prohibit the exchange of nutrition with the sea floor.

To conclude, these characteristics pave the way to stabilize biomolecular and cellular structures (e.g., DesMarais et al., 2002), such that liquid water is vital for all complex lifeforms that we know. No other known molecule matches all these capabilities when it comes to unique properties that support life. The essential dependence on water governs all known lifeforms and numerous large organisms that inhabit Earth's oceans.

### 2.1.2 Oxygen

Besides $H_2O$ molecules, dioxygen ($O_2$) in oceans and the atmosphere is deeply related to the origin, functionality, diversity, and size of lifeforms (e.g., Catling et al. 2005; Lyons et al., 2014; Ward et al., 2019; Balbi and Frank, 2023). If there are evolutionary conditions in which CALs evolve from bacteria to eukaryotic lifeforms such as mammals, including technological civilizations that require large complex brains, size considerations of such lifeforms are important (Catling et al., 2005; Balbi and Frank, 2023 and references therein). Before the GOE that occurred on Earth around 2.3 Gyr ago (e.g., Lyons et al., 2014), physical and chemical evidence for eukaryotes with complex cells was absent in the rock records, and anoxygenic photosynthetic bacteria initially dominated over cyanobacteria (e.g., Olejarz et al., 2021). As discussed in detail by Catling (2005), $O_2$ which plays an essential biochemical role in large complex lifeforms, is the only known high-potential oxidant that is sufficiently stable to accumulate in the atmospheres of planets. Metabolisms in aerobic environments are up to about ten times more energy-effective compared to anaerobic lifeforms for certain food intake (Catling et al., 2005; Balbi and Frank, 2023). About 90% of $O_2$ in mammals serves as a terminal electron acceptor[3] in the energy production of aerobic metabolism (Chapman and Schopf, 1983; Rolfe and Brown, 1997; Catling et al. 2005). Furthermore, metabolic oxidation of organics by $O_2$ produces much more free energy than any other plausible respiratory process (e.g., Towe, 1981; Catling et al., 2005; and references therein; Ward et al., 2019).

---

[3] A terminal electron acceptor in biology corresponds to a last compound that receives an electron in an electron-related transport chain, like oxygen during cellular respiration.

However, one should note that chlorine and fluorine are more energetic oxidants per electrons compared to $O_2$, but these species are much less abundant compared to oxygen (Balbi and Frank, 2023). While $O_2$ is the 3$^{rd}$ element, fluorine is the 24$^{th}$ and chlorine is the 23$^{rd}$ abundant element in the Universe. Compared to $O_2$ both alternative elements are rare trace elements. Budisa et al. (2014) argued that the absence of the involvement of fluorine in Earth's biochemistry is due to the low presence of fluorine under early Earth conditions. Although such a scenario might be different on another terrestrial planet, besides their low abundance, these alternative elements have undesired side effects in organic reactions at temperatures that are comfortable for life as we know it. Because of these qualities that are bad for CALs, one can conclude that $O_2$ will be the most energetic source that is available for CALs in planetary environments.

Another role of $O_2$ can be observed in fossil records that reveal a dependence of the growing potential of organisms on the evolution of the $O_2$ partial surface pressure (Raff and Raff, 1970; Runnegar, 1982; 1991; Catling et al. 2005; Ward et al., 2019). According to Catling et al. (2005), an estimate of the maximum size limit of complex lifeforms can be obtained by assuming that the metabolism of the innermost cell consumes oxygen at the rate of $O_2$ supply by diffusion such that the inner $O_2$ partial pressure drops to zero. Therefore, an environment with $O_2$ as atmospheric bulk gas provides a large feasible energy source, and one can expect that atmospheric $O_2$ is an essential molecule for the high energy demands of large aerobic complex lifeforms with sizes $\geq 1$ cm outside the Solar System (Catling et al. 2005).

According to the already mentioned study by Catling et al. (2005), an $O_2$ partial pressure on planetary surfaces of $\approx 10$ mbar is necessary for aerobic complex life to form sizes of $\approx 1$ mm. Crucially, a minimum limit on the $O_2$ partial pressure of $\gtrsim 10 - 100$ mbar is needed to exceed a certain threshold size of about 1 cm above which circulatory physiology for supplying oxygen to the entire body, for example through blood circulation, is physically possible; such a threshold enables the evolution of larger aerobic lifeforms that are not diffusion-limited. Intelligent extraterrestrial lifeforms that develop to a technological level to beam radio signals into space as suggested by the SETI research community would need a planet where the atmospheric $O_2$ surface pressure is not too low or should have evolved on a planet with atmospheric oxygen levels that lies within the so-called controllable fire zone (see Figure 1) (Knoll and Papagiannis, 1985; Catling et al., 2005; Judson, 2017; Balbi and Frank, 2023). Here, we note that the question of whether the origin of animals on Earth was triggered by a rise in atmospheric oxygen as well as the lower respiration limits of basal animals is hotly debated (see, e.g., Mills and Canfield 2014; Lyons et al. 2021; Sperling et al. 2022). However, the answer to these research questions does not affect the $O_2$ size threshold as discussed by Catling et al. (2005) and above.

On the other side of the pressure scale, it is known that a too-high $O_2$ partial pressure causes problems for mammals (Binger et al., 1927; Clark and Lambertsen, 1971; Thomson and Paton, 2014). Animal experiments have indicated a variation in tolerance similar to that found in central nervous system toxicity between different species. For $O_2$ partial pressure levels < 0.3 bar, humans are not harmed by $O_2$ at all, but for > 0.5 bar pulmonary toxicity symptoms set in (*e.g.*, Clark and Lambertsen, 1971); $O_2$ partial pressures of 1 bar and more cause irreversible damage (Bitterman and Bitterman, 2006). In addition, also $O_3$ can be toxic to animal-like life in high concentrations (> 40 ppb) at a planet's surface (Cooke et al., 2024), even though it is an important UV absorber at high altitudes.

But the available number of strong oxidants also affects the living environment chemically; most importantly, the oxygen surface mixing ratio controls fires (e.g., Watson et al., 1978; Watson and Lovelock 2013; Lenton and Watson 2000; Belcher et al. 2010a; 2010b; Belcher et al., 2021). If, for example, the atmospheric $O_2$ mixing ratio is too low, fires cannot burn because they are cooled faster than the oxidant is supplied, while if the atmospheric $O_2$ mixing ratio is too high, fires are difficult to put out because convection supplies the oxidant faster than it cools the fire.

An extraterrestrial intelligent civilization that progressed towards technology cannot exist in an environment where fires cannot be controlled and common organic materials start accidentally to burn (Swords, 1995; Balbi and Frank, 2023; Stern and Gerya, 2024). Controlled fire means that one can keep the fire going, but it will not rage out of control (Cloud, 1989). Experiments for which paper was used as fuel indicated that the lower limit of atmospheric $O_2$ mixing ratio for combustion in Earth's atmosphere occurs at $\approx 0.16$, while the probability of fire ignition can be assured at atmospheric $O_2$ mixing ratios $> 0.2$ (Belcher et al., 2008; Belcher et al., 2010).

Figure 1 illustrates the atmospheric $O_2$ mixing ratio related to fire zones and compares it with the $O_2$ surface partial pressure range inhabited by humans on present Earth (dashed lines). The latter corresponds to a constant atmospheric $O_2$ mixing ratio of $\approx 0.27$ and is clearly within the controllable fire zone (relevant for SETI, etc.).

Natural feedback might maintain atmospheric $O_2$ mixing ratios within the controllable fire zone (Lenton and Watson, 2002; Balbi and Frank, 2023). According to Wildman et al. (2004), depending on the moisture content atmospheric $O_2$ mixing ratios, $\geq 0.3$ might allow for the persistence of plants, but values of $\approx 0.35$ represent a final upper boundary for the existence of vegetation (Chaloner, 1989). For a 1 bar atmosphere, this zone lies within surface partial pressure values of $\approx 100 - 240$ mbar (Swords, 1995; Belcher et al., 2021; Balbi and Frank, 2023). On Earth, the so-called "oxygenation time", that is, the time needed for the $O_2$ partial pressure to reach values sufficiently high for the evolution of large aerobic complex lifeforms took around 3.9 Gyr (Catling et al. 2005 and references therein). Thus, the evolution of complex life as a vital step toward intelligent life is strongly dependent on the "oxygenation time" of a planet and, therefore, on sources and sinks for $O_2$ to accumulate and/or remain in an atmosphere over geological timescales.

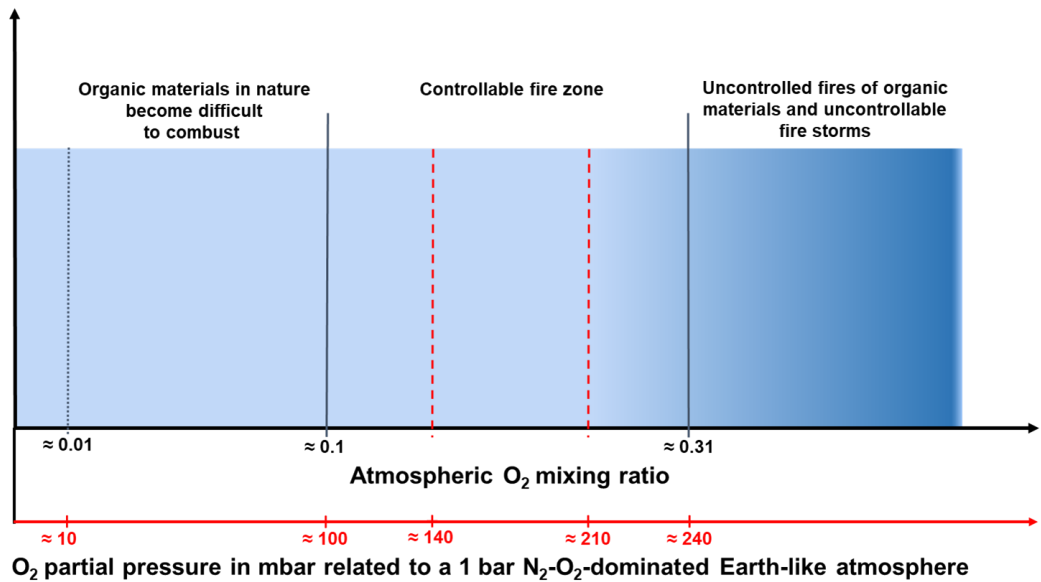

**Figure 1:** Illustration of atmospheric $O_2/N_2$ mixing ratios related to different fire zones and their corresponding partial $O_2$ surface pressure values in a 1 bar $N_2$-$O_2$-dominated Earth-like atmosphere. Below the mixing ratio of $\leq 0.1$ ($\leq 100$ mbar $O_2$ at present Earth) organic materials become difficult to combust, while at an atmospheric mixing ratio $\geq 0.31$ ($\geq 240$ mbar at present Earth) wildfires cannot be handled anymore (Cloud, 1989; Belcher et al. 2010a). The dashed lines show the partial $O_2$ surface pressures at the highest human settlements ($\approx 140$ mbar) and at sea level ($\approx 210$ mbar), which corresponds to an $O_2 / N_2$ mixing ratio of $\approx 0.27$.

Moreover, atmospheric $O_2$ and oxygen production through photosynthesis is also coupled with phosphorus, which is the limiting nutrient in oceanic environments over geological timescales (Tyrrell 1999; Karl 2014; Duhamel et al., 2021; Olejarz et al., 2021). The content of P derived from Earth's igneous rock record was found to correlate with its atmospheric $O_2$ history (Cox et al., 2018). On one hand, this is a strong indication that rising levels of $O_2$ during and after the GOE are coupled with the increased production of free oxygen through photosynthesis. On the other hand, this shows that there has to be a certain amount of phosphorus available – first in the mantle melts, later within the biosphere – to achieve such high biological productivity that sinks by burial on continental shelves and deep-sea sediments through, for example $S^{2-}$ or $Fe^{2+}$, are outpaced (Cox et al., 2018). From these findings, one can conclude that not only the initial nutrient inventory of a planet but also the speed of mantle cooling, which is also related to the existence of working tectonics, affect the biological evolution and, consequently, the atmospheric $O_2$ evolution of any EH.

### 2.1.3 Carbon dioxide

As briefly mentioned in Sect. 2, aerobic complex animal lifeforms are also affected by the amount of CO and $CO_2$ in an atmosphere (Schwieterman et al., 2019). On present Earth, CO is mainly produced by photolysis of $CO_2$ molecules and biological activity at the surface. High $CO_2$ partial pressure values impose crucial physiological stress across molecular, cellular, and organismal scales on complex organisms (e.g., Pörtner et al., 2004; Azzam et al., 2010; Wittmann and Pörtner, 2013). According to various studies, physiological stress on CALs caused by elevated $CO_2$ values during Earth's history may have been a factor in major mass extinctions (Clarksen et al., 2015; Knoll et al., 2007; Knoll, 2014). Recently, possible limitations for complex lifeforms due to atmospheric $CO_2$ and related CO concentrations were found to prevent the habitability for aerobic complex life when $CO_2$ concentrations exceed ~100 mbar (Schwieterman et al., 2019, Ramirez, 2020). Experiments with rhesus monkeys showed a possible acclimation up to $\approx 60$ mbar whereby sheep start to experience discomfort when the $CO_2$ reaches $\approx 80$ mbar and tissue damage or death at $\approx 120$ mbar and at $\approx 160$ mbar, respectively (see e.g.,

Knowlton et al., 1969; Hoover et al., 1970; Stinson et al., 1971; Ramirez 2020). It is also known that newborn mammals react more sensitively to the $CO_2$ partial pressure such that their upper $CO_2$ pressure limit is rather closer to $\approx$100 mbar (Kantores et al., 2006; Ramirez, 2020) in a 1 bar atmosphere. If the atmospheric $CO_2$ level falls below 0.15 mbar, most plants and, hence, all animals, including insects and other invertebrates that depend on plants for their survival, will die. The limitation related to atmospheric $CO_2$ for plants is related to photosynthesis, which needs a certain $CO_2$ concentration to ensure its functioning (*e.g.*, Lovelock and Whitfield, 1982). The main process of photosynthesis, which produces a three-carbon compound by the so-called Calvin cycle[4] is called C3 photosynthesis, while C4 photosynthesis belongs to a photosynthesis process that produces a four-carbon compound, which splits into a three-carbon compound for the Calvin cycle (Silverstein, 2008).

In a 1 bar atmosphere at a $CO_2$ partial pressure of $\approx$ 0.20 mbar, plants that use the C3-photosynthesis show a significant decrease in biomass production and no C3-plants survive at $CO_2$-partial pressure values of $\approx$ 0.05 mbar, which corresponds to the $CO_2$ amount at which a plant's photosynthesis rate equals its rate of respiration (e.g., Gerhart and Ward, 2010 and references therein). For plants that operate on the C4-photosynthesis process, the atmospheric $CO_2$ limit is reached at $\approx$ 0.01 mbar. The atmospheric $CO_2$ values are also linked to an increase in the luminosity of the host star, which leads to enhanced weathering rates on EHs and, hence, a decrease of atmospheric $CO_2$ over geological timeframes (Caldeira and Kasting, 1992, Franck et al., 2002).

Earth's present $CO_2$ atmospheric partial pressure is at $\approx$ 0.4 mbar, which lies closer to the lower, as opposed to the upper, limit. Thus, one can summarize that under consideration of the limitations for aerobic complex lifeforms related to atmospheric $CO_2$-CO surface partial pressure values and concentrations, the HZCL has to be preferred over the HZ, which reduces habitable orbits around Sun-like stars to only $\approx 21 - 32$ % (Schwieterman et al., 2019) or $28 - 43$ % (Ramirez, 2020) of the initial value, depending on the assumed threshold for the $CO_2$ partial pressure (see Scherf *et al.* (2024, this issue) for further details).

---

[4] In photosynthesis, the Calvin cycle corresponds to the chemical reactions that convert atmospheric $CO_2$ and hydrogen-carrying compounds into glucose. The Calvin cycle occurs in all photosynthetic eukaryotes and most photosynthetic bacteria.

For 1 bar atmospheres, together with the previously discussed atmospheric $O_2$ partial pressure values of $\geq 100$ mbar required for the development of m-sized complex lifeforms, the coexistence with the right amount of these atmospheric species as a bulk gas has important implications for the search for potential $\eta_{EH}$-planets as well as their frequency and, hence, for the search of extraterrestrial (intelligent) lifeforms in general.

### 2.1.4 Nitrogen

The most abundant species in Earth's bulk atmosphere is $N_2$. Nitrogen is one of the building blocks of life (incorporated in amino acids, proteins, and DNA), and it is therefore essential for all plants and animals.

All aerobic complex lifeforms require nitrogen in a reactive form (often referred to as "fixed"), whose major source on Earth is the biological fixation of atmospheric $N_2$ (*e.g.*, Galloway, 2003; Fowler et al, 2013). The labile equilibrium between nitrogen-related nutrient availability and lifeforms can be simply condensed: In most ecosystems, a low supply of fixed nitrogen – besides phosphorus – is considered a limitation for flora (autotroph) growth, but also a very high sustenance can lead to local extinction, for instance in so-called marine dead zones where whole ecosystems die out (e.g., Howarth and Marino, 2006; Elser et al., 2007). Klingler et al. (1989) studied the growth rates of two bacteria species capable of fixing nitrogen called "Azotobacter vinelandii" and "Azomonas agilis" under various nitrogen partial pressures from zero to 780 mbar.

These authors found that, below ≈ 400 mbar the biomass, cell number and growth rate decreased along with decreasing nitrogen partial pressure, but both microorganisms were capable of growing until the partial pressure reached ≈5 mbar; absolutely no microbial growth was observed below 1 mbar. These findings indicate a lower limit of 1-5 mbar nitrogen partial pressure for aerobic microorganisms to perform biological fixation, at least if these constraints are similar for other nitrogen-fixing species (Klingler et al., 1989).

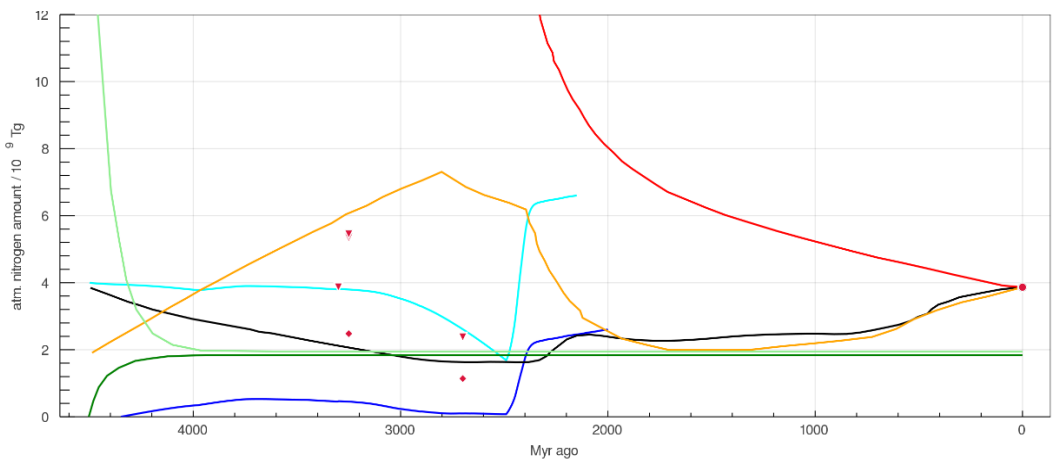

**Figure 2:** Comparison between different model runs from different studies for Earth's nitrogen partial pressure throughout time. The red dot shows the present Earth's $N_2$. The red and yellow lines show results of Johnson and Goldblatt (2018) referred to as "normal" and "normal run but with low starting atmospheric mass". The two green lines represent the "high fixation" scenarios by Laneuville et al. (2018) with different initial $N_2$ levels (these runs are only for lifeless worlds and are, therefore, not affected by the Great Oxidation Event of around 2300 Myr ago). The black line indicates the "heat flow" model by Stüeken et al. (2016). The turquoise and blue lines show a model run by Sproß et al. (2021) with different initial $N_2$ atmospheric inventories. Atmospheric $N_2$ constraints by Marty et al. (2013) for 3250 Myr ago, Avice et al. (2018) for 3300 Myr ago, and Som et al. (2016) for 2700 Myr ago are drawn as red diamonds (best value) and triangles (maximum).

According to Bennett and Rostain (2003) and Ramirez (2020), serious intoxication through $N_2$ occurs at ≈3 bar for a 78 % $N_2$-dominated atmosphere by assuming that $O_2$ has a similar narcotic toxicity as $N_2$. Ramirez (2020) investigated for the first time $N_2$ respiration limits for complex lifeforms inside the HZCL and found an upper limit of around 2 bars for newborn and adult mammals.

It should also be noted that high nitrogen partial pressures would also contribute to the greenhouse effect due to increased $N_2$-$N_2$ collision-induced absorption and a decrease in outgoing infrared flux (*e.g.*, Johnson and Goldblatt, 2018; Lammer et al., 2019; Ramirez, 2020). Further, it must be mentioned that a variety of abiotic processes such as aurorae, lightning, impactors, volcanoes, and more can convert atmospheric $N_2$ to fixed nitrogen even in $CO_2$ atmospheres (Lammer et al., 2019).

In Figure 2, the influence of life on the $N_2$ partial pressure based on modeling of the early Earth's reducing environment is shown. Biological nitrogen fixation (BNF) is so intense that it completely depletes atmospheric nitrogen within very short periods, if an effective return flux is missing (Lovelock 2001; Lammer et al. 2018, 2019; Sproß, 2019; Sproß et al. 2021; Stüeken et al., 2016). During the GOE transition, denitrification starts to contribute to and elevate atmospheric $N_2$ levels possibly strongly through releasing previously fixed and sedimented nitrogen back into the atmosphere (Stüeken et al. 2016a, 2020; Sproß et al. 2021).

Most nitrogen in Earth's lithosphere originated from biogeochemical processing and fixation into organic matter and the related transfer into low-temperature mineral phases like clay minerals (*e.g.*, Halama and Bebout, 2021; Busigny and Bebout, 2013).

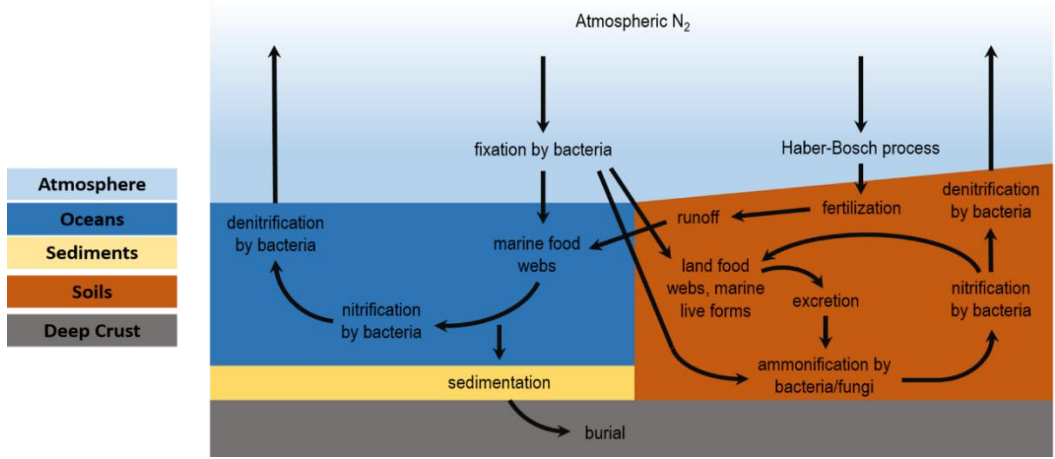

**Figure 3:** Illustration of Earth's nitrogen cycle showing lifeforms that are fixing nitrogen into the surface and bacteria that release nitrogen from the surface via denitrification. Since the buildup of atmospheric $O_2$ during the GOE at ≈2.3 Gyr, nitrogen was not only fixed by bacteria into the soil but also released into the atmosphere by heterotrophic bacteria. Higher atmospheric $N_2$ surface partial pressures of ≈800 mbar is also relevant for the protection against high energy cosmic ray particles for large aerobic complex surface lifeforms.

Nitrogen is then stored in silicate phases as $NH_4^+$ from where it can be retained efficiently in rocks that experience melting or partial melting phases (Bebout et al., 2013; Yang et al., 2017; Mikhail et al., 2017). In general, the nitrogen cycle shown in Figure 3 can be divided into five main stages:

- Nitrogen fixation of atmospheric $N_2$ to $NH_3/NH_4^+$ (biotic) or $NO_2^-$, $NO_3^-$ (abiotic);

- Oxidation ("nitrification") of $NH_3/NH_4^+$ to $NO_2^-$, $NO_3^-$;

- Assimilation and incorporation of $NH_3/NH_4^+$ and $NO_2^-$, $NO_3$ into biological tissue;

- Ammonification that transfers organic nitrogen compounds back to $NH_3/NH_4^+$;

- Conversion of $NO_2^-$, $NO_3^-$ and release of $N_2$ ("denitrification") into the atmosphere.

One can conclude that the conditions for maintaining an $N_2$- and $O_2$-based biosphere within the HZCL are a complex subject. Figure 4 illustrates a summary of the above-discussed restrictions regarding required atmospheric environments where CALs can exist.

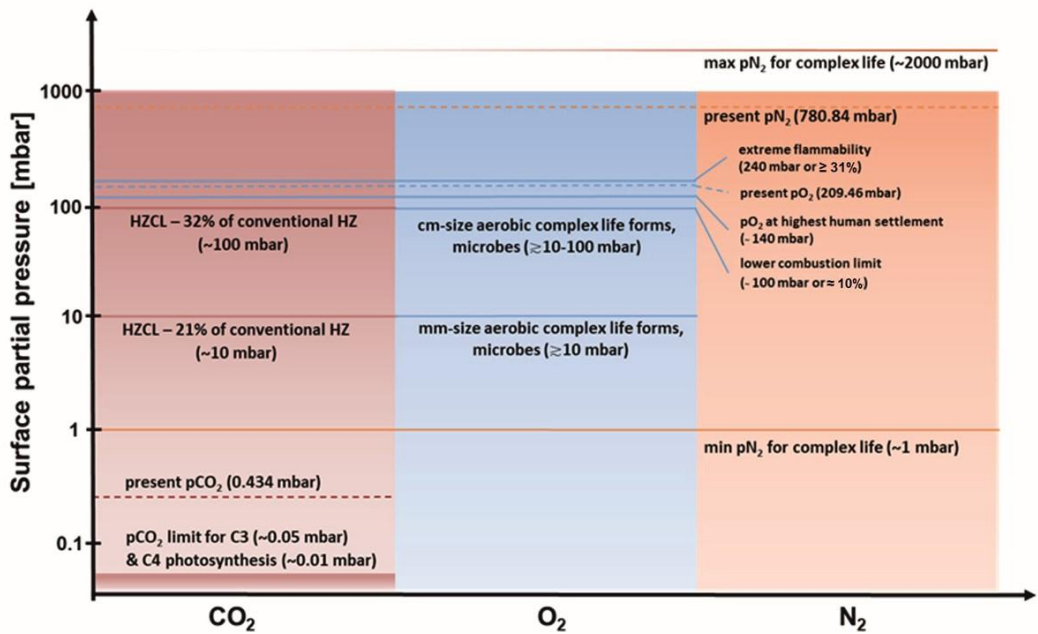

**Figure 4:** Surface partial pressures for a 1 bar total atmospheric pressure that constrain the environments in which CALs can develop and exist, as described in the text. Pressure limits according to Schwieterman et al. (2019) for HZCL-$CO_2$ limits; Gerhart and Ward (2010), and references therein; Sword (1995), and references therein, for $O_2$ flammability and human limits; Catling et al. (2005) for $O_2$ complex life limits; Ramirez (2020) for $N_2$ limits.

A big issue of the early Earth's reduced nitrogen cycle before the GOE is a likely low efficiency/availability of denitrification in oceans in environments, which are relatively poor in oxygen (*e.g.*, Zerkle et al., 2017; Lammer et al., 2019). Assuming similar fixation rates through abiotic processes as observed today – for instance, through lightning that can form HCN and $NH_x$ in early the Earth's reduced atmosphere – the oceans/soils receive a lot of nitrogen from the atmosphere; but there is no clear pathway to recycle $N_2$ back to the atmosphere. This would lead to a very unstable atmospheric pressure concerning a depletion of nitrogen. Especially if bacteria capable of nitrification contribute to the fixation, the time scale of a complete loss of 1 bar nitrogen can be as little as about 15 Myr (Jacob, 1999). Therefore, either the early Earth's nitrogen partial pressure was rather low or there was an efficient recycling process that was replaced by denitrification when the atmosphere became oxidized.

The geologic history of seawater acidity, which is critical for understanding seawater biochemistry, is relatively poorly constrained. Knauth (2005) inferred paleo-surface temperatures during the Archean eon from the analysis of oxygen isotope data from Cenozoic sediments of $\approx 328 - 358$ K ($\approx 55 - 85$ ºC).

These temperatures could have been caused by much higher atmospheric partial $CO_2$ surface pressures, of which the latter is in agreement with estimates from paleosols (Kanzaki and Murakami 2015). In such environmental conditions, one would obtain a relatively high acidity of the early oceans. Here, a by-product during oxygen production by cyanobacteria is $Fe_2O_3$, which was sedimented in high amounts throughout Earth's history; estimating the ferrous iron concentration in the early oceans based on the amount of deposited iron ores leads to about 100 times the modern level (Deamer et al., 2019, chap. 1). Since Ranjan et al. (2019) researched the possibilities of a non-biotic way of releasing nitrogen back into the atmosphere in ferrous iron-rich water, this process remains as a possible solution for the early Earth's nitrogen cycle closure. This might be only possible, however, in areas where photoferrotrophic bacteria supply the required nitrogen (Stüeken et al., 2020). In addition, Halevy and Bachan (2017) only inferred a slight decrease in acidity from the early Archean to the Phanerozoic.

Alternatively, Kasting et al. (1993) proposed nitrogen recycling based on the energy of mid-ocean ridge volcanism, which has never been investigated in more depth. In addition, the contribution of anammox, which might have already been developed before the GOE, is not clear (Brunner et al., 2013; Stüeken et al., 2016). Before the GOE during the Archean eon, a different form of the nitrogen cycle was in place, wherein an efficient release of nitrogen via denitrification was not possible (compare Figure 4a in Lammer et al., 2019). Such recycling of previously fixed nitrogen through denitrification (and/or anammox) is only possible because of oxidized nitrogen compounds and, therefore, requires an active aerobic part of the nitrogen cycle.

As one can see from these brief discussion**s,** Earth's present atmosphere is a result of biological activity, at least since the GOE. Thus, if one assumes that similar biosphere-atmosphere interaction processes take place at planets on which life originated, one can expect that such an exo-biosphere will also have a major influence on the environmental conditions of the particular planet either on a regional or a global scale.

On the other hand, one should keep in mind that the interaction between a biosphere and atmospheric constituents produced or maintained by microbial life are very complex and interconnected systems that can, but likely not exactly, develop similar to those on Earth. The emission and the mixing of atmospheric constituents related to lifeforms will influence atmospheric chemistry, the physical climate through photochemical processes, and the planet's radiation budget (e.g., Scholes et al., 2008). Coevolutionary feedback may lead to self-regulating processes, so-called homeostasis, by which organisms tend to maintain stability while adjusting to conditions that are best for their survival. If such processes are successful, life continues to evolve; if they are unsuccessful, it results in a disaster or death of the organisms.

In Earth's case of the GOE, for example, many microbes might have become extinct because atmospheric $O_2$ was toxic to them (*e.g.*, Olejarz et al., 2021). As another example, nitrogen fixation may have fixed almost the entire $N_2$ from the atmosphere before the GOE (Gebauer et al., 2020; Sproß et al., 2021); this could have also led to the mass extinction of some lifeforms.

Humans, and potentially some extraterrestrial technological civilizations as well, certainly do not live in homeostasis with the planet, due to climate-gas flux interactions, deforestation, agricultural technologies, etc. One should keep in mind that such anthropogenic influences can lead to significant alterations in the sources and sinks of trace gases, which can result in disastrous planetary environments that could affect the lifetime of hypothetical ETI civilizations.

To conclude, one can expect that the interaction between life and planets/atmospheres does not necessarily have to lead to a balance between a biosphere and atmosphere, as it was during the past billion years in the case of Earth.

## 3.  Deriving a formula for Earth-like Habitats

As discussed in Sect. 1, one of the main goals of exoplanet research is to determine $\eta_{\oplus}$, the frequency of hypothetical Earth-like exoplanets. So far, all previous studies that estimated the galactic number of Earth-like planets, or even the frequency of ETIs (*e.g.*, Irwin et al., 2014; West and Conselice, 2020; Bryson et al., 2021; Cai et al., 2021), neglected astrophysical environmental limits for $N_2$-$O_2$-dominated atmospheres being stable against atmospheric escape as induced by short-wavelength radiation (X-ray and extreme ultraviolet radiation, *i.e.*, XUV) and against the growing bolometric luminosity of their host stars. Moreover, several other important geophysical processes that are related to the evolution of Earth-like Habitats, such as the importance of continents, oceans, and plate tectonics relevant for carbon- and nitrogen cycles to operate smoothly and continuously over billions of years (Stern, 2016; Zaffos et al., 2017; O'Neill et al., 2021; Unterborn et al., 2015; 2022; Stern and Gerya, 2024) are also neglected in these studies.

Besides G stars, planets around the more numerous K and even M-dwarfs are often assumed to generally be habitable. Westby and Concelice (2020), for instance, estimated that there should be at least $36^{+175}_{-32}$ or even $928^{+1980}_{-818}$ ETI civilizations within our Galaxy that may communicate with radio signals. However, these authors suggested that civilizations would mostly send their signals from an M-star system. When spread uniformly throughout the Milky Way, the closest civilization in this study would populate a terrestrial planet in a low-mass M-dwarf star system at a distance of $17000^{+33600}_{-10000}$ light-years for their strictest case.

As indicated by several aeronomical studies for almost two decades (e.g., Khodachenko et al. 2007; Lammer, 2007; 2009; 2019; Tian et al., 2008a; 2008b; Airapetian et al., 2017; Dong et al., 2017; Johnstone et al., 2021), terrestrial planets in M-star systems are most likely not suited for the evolution of $N_2$-$O_2$-dominated atmospheres where CALs can, in principle, develop. Given that, we propose a formula that estimates the maximum number of Earth-like Habitats, $N_{EH,}$ on which CALs as defined in Sect. 1 can evolve.

### 3.1 Estimating the maximum occurrence rate of Earth-like Habitats

Our formula is based on the present number of stars, $N_*$, in the Galaxy (or any other set of stars) and on certain necessary habitability requirements that have to be met for a planet to evolve into an Earth-like habitat. Whether a planet evolves into an EH is dependent on $N_*$ itself but also on the fraction, $\eta_*$, of $N_*$ that can host planets with $N_2$-$O_2$-dominated atmospheres and on the fraction of rocky planets within the HZCL, $\eta_{EH}$, on which such atmospheres can develop. Based on this, we can write a formulation as follows, that is,

$$N_{EH} \leq N_* \times \eta_* \times \eta_{EH}. \tag{2}$$

Here, it should be noted that $N_{\mathrm{EH}}$ will always be less-equal to the estimated number as derived from the right-hand-side of this equation. The reason for this is related to the requirements that feed into $\eta_\star$ and $\eta_{\mathrm{EH}}$; since we cannot scientifically quantify each of the relevant criteria needed for EHs to evolve, we can only consider the ones that are known and already scientifically quantifiable. Any other relevant parameter will be ignored, that is, set to 1, until new scientific data will allow us to evaluate and implement it into our formulation.

These requirements can, for simplicity, be further divided into several related clusters; for $\eta_\star$, we can then write,

$$\eta_* \leq A_{\mathrm{GHZ}}\left(\textstyle\prod_{i=1}^{n} \alpha_{\mathrm{GHZ}}^{i}\right) \times A_{\mathrm{at}}\left(\textstyle\prod_{i=1}^{n} \alpha_{\mathrm{at}}^{i}\right). \qquad (3)$$

Here, $A_{\mathrm{GHZ}}\left(\prod_{i=1}^{n} \alpha_{\mathrm{GHZ}}^{i}\right)$[5] is the fraction of stars that reside within the so-called galactic habitable zone, famously abbreviated as GHZ (Lineweaver et al., 2004; Kaib, 2018; Scherf *et al.* (2024, this issue), and references therein), which again depends on various necessary criteria, $\alpha_{\mathrm{GHZ}}^{i}$. These, for example, include sterilization events such as supernovae, $\alpha_{\mathrm{GHZ}}^{\mathrm{SN}}$, and the metallicity distribution, $\alpha_{\mathrm{GHZ}}^{\mathrm{met}}$, within the Galaxy. Further, $A_{\mathrm{at}}\left(\prod_{i=1}^{n} \alpha_{\mathrm{at}}^{i}\right)$ denotes the fraction of stars that permit the long-term sustainable existence of an $N_2$-$O_2$-dominated atmosphere within the partial pressure values discussed in Section 2. The parameter $A_{\mathrm{at}}$ can, among others, scientifically be constrained by the young star's XUV flux, $\alpha_{\mathrm{at}}^{\mathrm{XUV}}$, which should be below a certain threshold value to avoid complete escape of such an atmosphere (see, *e.g.*, Tian et al. 2008a; 2008b; Lammer et al. 2008; 2018; Lichtenegger et al., 2010; Johnstone et al., 2021; Gebauer et al., 2020; Kislyakova et al., 2020; Stüeken et al., 2020). A second restricting boundary, $\alpha_{\mathrm{at}}^{S_{\mathrm{eff}}}$, is related to the host star's bolometric luminosity evolution and the related stellar effective surface flux, $S_{\text{eff}}$, a planet in the HZCL experiences, which terminates planetary habitability due to the evaporation of a planet's liquid water oceans or through enhanced $CO_2$-weathering rates (Lovelock and Whitfield, 1982; Caldeira and Kasting, 1992; Franck et al., 2002), as mentioned in Section 2. Both requirement terms, $A_{\text{GHZ}}$ and $A_{\text{at}}$, and the factors that feed into them will be discussed and evaluated in detail in Scherf *et al.* (2024, this issue).

---

[5]The letters $A$ and $\alpha$ were chosen based on the Greek word for star, i.e., αστερι.

For $\eta_{\text{EH}}$, we can derive a similar formulation, that is,

$$\eta_{\text{EH}} \leq \beta_{\text{HZCL}} \times B_{\text{pc}}\left(\prod_{i=1}^{n} \beta_{\text{pc}}^{i}\right) \times B_{\text{env}}\left(\prod_{i=1}^{n} \beta_{\text{env}}^{i}\right) \times B_{\text{life}}\left(\prod_{i=1}^{n} \beta_{\text{life}}^{i}\right). \quad (4)$$

Here, the first parameter that feeds into $\eta_{\text{EH}}$ is the occurrence rate, $\beta_{\text{HZCL}}$[6], of rocky exoplanets within the HZCL, a factor that is almost equivalent to $\eta_{\oplus}$ but is scaled to the habitable zone for complex life. In addition, three clustered requirements feed into $\eta_{\text{EH}}$ which indicates that $\eta_{\text{EH}}$ will be smaller than $\eta_{\oplus}$. The first of these, that is, $B_{\text{pc}}\left(\prod_{i=1}^{n} \beta_{\text{pc}}^{i}\right)$, denotes the fraction of planets that form and end up with a planetary composition and mineralogy that allows for the subsequent evolution of aerobic lifeforms. This parameter, for instance, includes $\beta_{\text{pc}}^{H_2O}$, the necessary factor of accreting a suitable amount of water (*e.g.*, Stern, 2016; Zaffos et al., 2017; Stern and Gerya, 2024). A related criterion that feeds into $B_{\text{pc}}$ is the fraction of planets that did not end up with an accreted primordial $H_2$-He-dominated atmosphere, which can be depicted as $\beta_{\text{pc}}^{\text{acc}}$.

The second parameter in Eq. 4, $B_{\text{env}}\left(\prod_{i=1}^{n} \beta_{\text{env}}^{i}\right)$ clusters any requirements that are needed for the long-term environmental stability of Earth-like Habitats. These, for example, include $\beta_{\text{env}}^{\text{cycle}}$, the requirement of working carbon-silicate and nitrogen-cycles to be present at the planet over geological timescales. It may also

---

include debatable parameters such as $\beta_{\text{env}}^{\text{moon}}$, the fraction of planets that have large moons, and $\beta_{\text{env}}^{\text{mag}}$, the fraction of planets with a long-term stable magnetic field (see Scherf *et al.* (2024, this issue) for a discussion on their potential importance).

The last clustered parameter, $B_{\text{life}}\left(\prod_{i=1}^{n} \beta_{\text{life}}^{i}\right)$, includes any biological requirement that is needed for CALs to evolve on a planet such as the origin of life, $\beta_{\text{life}}^{\text{origin}}$. This cluster is important since without life no $N_2$-$O_2$-dominated atmosphere can likely evolve on a planet (see, *e.g.*, Lammer et al. 2019; Sproß et al. 2021). However, with the present state of scientific knowledge, this entire cluster may be set equal to 1 since the likelihood of the origin of life and of any other factors related to $B_{\text{life}}$ such as the appearance of photosynthesis are unknown. If we even set the parameters ($B_{\text{pc}}\left(\prod_{i=1}^{n} \beta_{\text{pc}}^{i}\right)$, $B_{\text{env}}\left(\prod_{i=1}^{n} \beta_{\text{env}}^{i}\right)$, and $B_{\text{life}}\left(\prod_{i=1}^{n} \beta_{\text{life}}^{i}\right)$) in Eq. 4 equal to 1 and substitute $\beta_{\text{HZCL}}$ with $\eta_{\oplus}$ then $\eta_{\text{EH}}$ will correspond to the conventional $\eta_{\oplus}$ parameter. This illustrates why $N_{\text{EH}}$ will be a strict maximum number due to setting any non-quantifiable requirement equal to 1.

Finally, we point out that some of the necessary requirements that feed into our formula could be correlated with each other. If some of them are positively or negatively correlated, the result will be either an underestimate (for a positive correlation) or an overestimate (for a negative correlation), since the probability of both requirements being valid together will be either higher or lower than the probability of both taken separately. There are obvious examples where requirements correlate, and we usually point them out in Scherf *et al.* (2024, this issue) when discussing them, at least as long as any correlations are known or suspected. The water requirement might be positively correlated with the potential requirement of plate tectonics (*e.g.*, Stern and Gerya 2024). Having plate tectonics may positively correlate with having a large moon. As an opposite example, there could be a negative correlation between having a large moon and the water requirement since a moon-forming giant impact could theoretically desiccate a protoplanet from its water resources. However, whether there is indeed a correlation between these specific requirements is not known.

Another class of correlation concerns whether only one combination of requirements is sufficient for the evolution of EHs or if there is more than one. While at least one combination must exist - otherwise Earth itself would not exist - it is unclear whether there could be more than one sufficient combinations, implying that not each requirement is necessary for every EH evolution. One could, for instance, hypothetically imagine that one EH has liquid water and plate tectonics (but no large moon) both of which induce a functioning C-Si cycle. In contrast, another EH has liquid water and a large moon (but no plate tectonics) both of which induce a functioning C-Si cycle through moon-induced tides.

In the following, we illustrate two specific examples that feed into $B_{\text{pc}}$ and $B_{\text{env}}$ are not scientifically quantifiable at present but will most likely be quantifiable within the next two decades due to atmospheric characterization of discovered rocky exoplanets by present and proposed ground- and space-based telescopes such as JWST, LIFE, HWO, and the ELT. Any further factors will be discussed in Scherf *et al.* (2024, this issue) of our study.

### 3.2 Example requirements that feed into $\eta_{\text{EH}}$ that are constrainable in the near future

Given the complexity of the topic, some of the required data in Eq. 4 will either remain unknown or inaccurate during the coming decades or can only be estimated by using theoretical models. The parameters $\beta_{\text{pc}}^{\text{acc}}$ and $\beta_{\text{env}}^{\text{cycles}}$, however, will allow statistical restrictions in Eq. 4 through upcoming observations of primordial $H_2$-He-dominated and/or $CO_2$-dominated atmospheres on low-mass terrestrial exoplanets within the next decade. The discovery of such atmospheres on these planets, even at orbital locations closer to the star than their corresponding habitable zones, will contribute to the statistics of these parameters. For instance, if a certain fraction of terrestrial planets in an orbital location comparable to Mercury in the solar system did not lose their primordial atmospheres, then one can expect that such planets would also not have gotten rid of accreted $H_2$-He-envelopes at orbital locations inside their habitable zones where atmospheric escape processes are less efficient. The observation of such atmospheres together with accretion models for terrestrial planets can then be used for fine-tuning parameters like $\beta_{pc}^{acc}$ in Eq. 4.

Similarly, the discovery of Mars- to Venus-like $CO_2$-dominated atmospheres around a certain fraction of low-mass terrestrial planets in or outside their habitable zones will also contribute to statistics related to the existence of stable C-silicate and N-cycles and would feed into the parameter $\beta_{env}^{cycles}$ of Eq. 4. Below, we examine these two parameters in more detail to inform that the observations of primordial $H_2$-He and/or $CO_2$-dominated atmospheres at low mass rocky exoplanets in the near future will contribute in better constraining $\eta_{EH}$.

### 3.2.1 Accretion of primordial atmospheres ($\boldsymbol{\beta_{pc}^{acc}}$)

One of the requirements $B_j$ that have to be met by any Earth-like Habitat is the initial planetary composition requirement, $B_{pc}(\prod_{i=1}^{n}\beta_{pc}^i)$, such that the number of certain elements like volatiles and radioactive elements accreted and sustained by a planet needs to be appropriate for complex aerobic life to originate and to evolve. This includes factors such as the accretion-related accumulation of primordial atmospheres from the gas disk $\beta_{pc}^{acc}$ and the right amount of water, $\beta_{pc}^{H2O}$. The first factor, $\beta_{pc}^{acc}$, that is often overlooked, is the accretion of $H_2$ and other gases such as He from the protoplanetary disk, as we will see in the subsequent paragraphs.

From the discovery of low-mass exoplanets, it is known that the largest discovered population to date has a mass/radius between those of Earth and Neptune (*e.g.*, Mullally et al., 2015) demonstrating that the variety of planets that exist in nature is much larger than previously thought. Although exoplanets with various densities/compositions certainly exist, this spread is tightly related to the existence of a large variety of low-mass planets that accreted within the gas disk and did not lose their accumulated primordial $H_2$-He-dominated atmospheres (e.g., Lammer et al., 2014; Owen et al., 2020).

Figure 5 shows examples of two accretion scenarios of Earth-like planets; one that accreted $0.6 M_\oplus$ within the lifetime of the protoplanetary gas disk, and a second that accreted $1.0 M_\oplus$ in the gas disk. For both scenarios, the disk lifetime was assumed to be 5 Myr (for details see, Erkaev et al., 2023). The loss of the $H_2$-dominated primordial atmospheres is related to the evolving stellar EUV-flux of the host star (*e.g.*, Tu et al., 2015). Here, both planets are exposed to highly (blue lines), moderately (green lines), and weakly active (red lines) young G-type stars. One can see that, in some cases, Earth accreted too much $H_2$ and a part of the accumulated primordial atmosphere would remain for more than 1.0 Gyr or even until the planet's lifetime. These Earth-mass planets, although orbiting inside the HZCL, end up as sub-Neptune's, and no complex aerobic lifeforms could evolve. From this brief discussion, one can expect myriads of scenarios that yield different Earth-size/mass planets with various primordial atmospheric masses and volatile abundances after they finish their accretion.

The role of the above-mentioned processes related to the evolution of Earth-like Habitats has so far been neglected, however. We expect that such disk-accretion-related processes as consolidated in the parameter $\beta_{\mathrm{pc}}^{\mathrm{acc}}$ within $B_{\mathrm{pc}}(\prod_{i=1}^{n} \beta_{\mathrm{pc}}^i)$ can further constrain the number of potential EHs on which CALs can evolve. The parameter $\beta_{\mathrm{pc}}^{acc}$, therefore, gives the fraction of terrestrial planets that will not end up with $H_2$- or He-dominated primordial atmospheres. This can be statistically fine-tuned in the near future when $H_2$/He-dominated atmospheres around Earth-mass exoplanets become observable. The determined $\beta_{\mathrm{pc}}^{\mathrm{acc}}$ values can then be refined with each further observation.

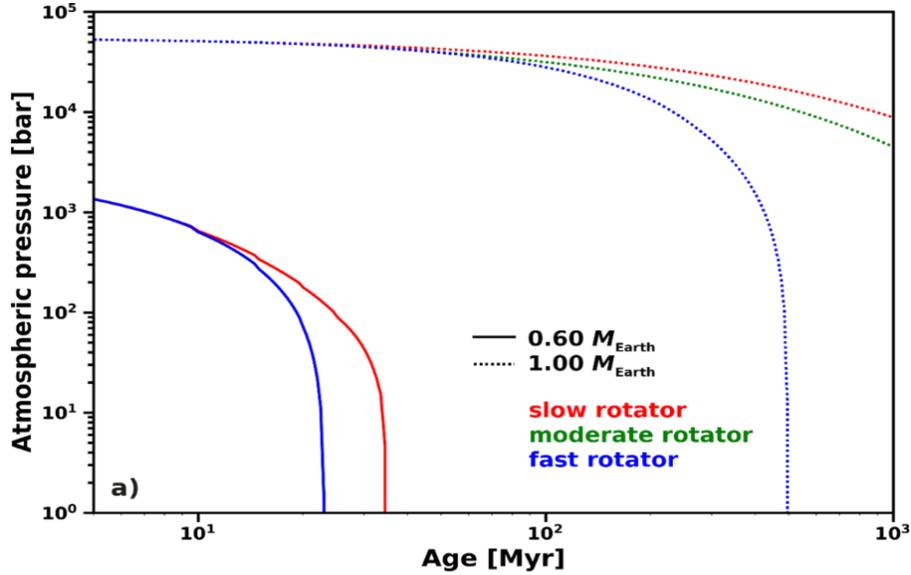

**Figure 5:** EUV-driven hydrodynamic escape of H₂-dominated primordial atmospheres for two different accretion and EUV flux scenarios. The solid lines correspond to a case where proto-Earth accreted $0.6M_\oplus$ within the protoplanetary gas disk; in such a case, the primordial H₂-envelope is lost within ~ 23 - 34 Myr, while the protoplanet accretes to its final mass of $1.0M_\oplus$ (see Erkaev et al., 2023, for details). Here, the EUV-activity evolution of the young Sun corresponds to a highly active (blue line) and weakly active (red line) G-type star, as described by Lammer et al. (2020a). The dotted lines correspond to an accretion scenario in which the protoplanet already accreted to its final mass of $1.0M_\oplus$ during the nebula phase. This scenario is exposed to highly (blue line), moderately (green line), and weakly active (red line) G-type stars. For the moderately and weakly active scenarios, the primordial atmosphere is not lost within 1.0 Gyr and may even remain during the whole lifetime of the planet.

Another important factor that is (partially) related to the accretion history of Earth-like planets and that resides within $B_{pc}(\prod_{i=1}^{n} \beta_{pc}^i)$ is $\beta_{pc}^{H2O}$, the fraction of potential EHs that possess the right amount of liquid water for both oceans and continents to exist. An Earth-size/mass planet is not necessarily a terrestrial "rocky" Earth-like planet. Planetary bodies that accrete slowly enough to avoid accumulating too much nebula gas but originate beyond the snow line should, for instance, be rich in volatile ices like $H_2O$ and $NH_3$ (*e.g.*, Kucher, 2003; Léger et al., 2004). There will be a fraction of such planets that can migrate inward due to interactions with the disk or other planets. According to Kucher (2003), such planets can mimic terrestrial planets inside the HZCL, but they can retain their volatiles and thick steam-like atmospheres for billions of years. One can expect that a factor like $\beta_{pc}^{H2O}$ can also be statistically evaluated in the future by the spectroscopic characterization of terrestrial exoplanet atmospheres inside the HZCL.

### 3.2.2 Long-term stable C-silicate and N-cycles ($\beta_{env}^{cycles}$)

For terrestrial planets that accreted the right mass within the right time to not end up with $H_2$-He-dominated primordial atmospheres, as mentioned above, it will be important to evolve long-term stable, well-working carbon-silicate and nitrogen-cycles for which, at Earth, several geological key processes are needed (see, *e.g.*, Halama and Bebout, 2021, and references therein; Stern and Gerya, 2024). For the evolution of these cycles, a terrestrial planet needs the right amount of heat production and water abundance for its thermal evolution (*e.g.*, O'Neill et al., 2020; Unterborn et al., 2015; 2022). Different initial amounts of radioactive heat-producing isotopes such as $^{40}K$, $^{235}U$, $^{238}U$, $^{232}Th$ with their corresponding half-life times of $t_{1/2;40K} \approx 1.25$ Gyr, $t_{1/2,235U} \approx 0.704$ Gyr, $t_{1/2,238U} \approx 4.468$ Gyr, and $t_{1/2,^{232}Th} = 14.05$ Gyr have important consequences for the thermal evolution of terrestrial planets (*e.g.*, Turcotte and Schubert, 2002; O'Neill et al., 2016; O'Neill et al., 2021 and references therein); these isotopes, therefore, determine to a great extent whether a terrestrial planet evolves to an EH or not (*e.g.*, Erkaev et al. 2023).

If present-day Earth's internal heat production was lower by a factor of 2, the mantle potential temperature would increase and convective stress levels would fall so that the initial mode of plate tectonics would be unstable to natural perturbations (Lenardic et al., 2011; O'Neill et al., 2021). As shown in Figure 6, different initial K, but also U, Th abundances, will therefore result in different heat budgets in planetary interiors, which has important implications for the thermal evolution of planetary interiors and global processes such as the generation of long-lived magnetic dynamos (Turcotte and Schubert, 2002; Murthy et al., 2003; Nimmo, 2007) and tectonics (*e.g.*, O'Neill et al., 2021).

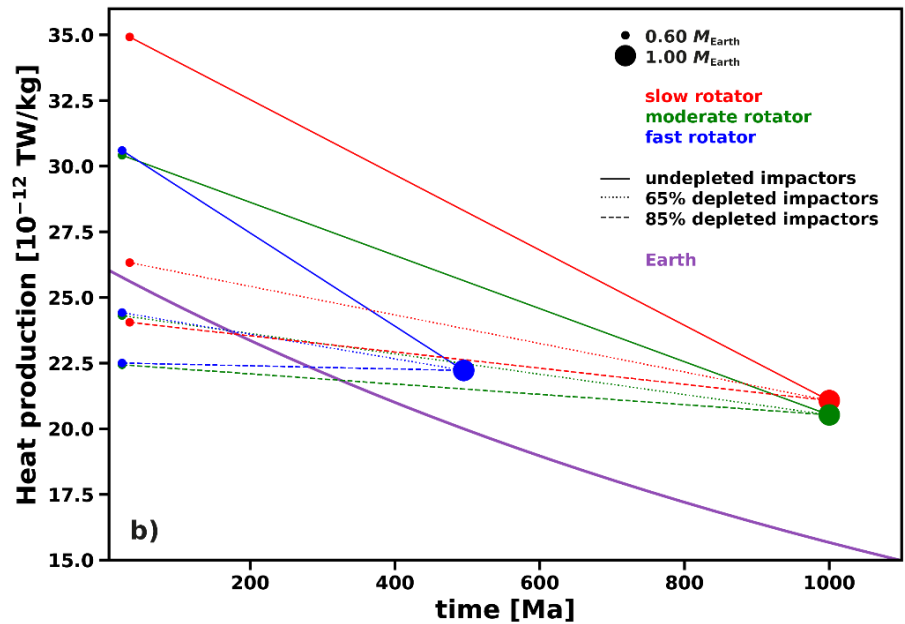

**Figure 6:** Depiction of the final radioactive heat production by $^{235}$U, $^{238}$U, $^{232}$Th and the remaining $^{40}$K for the same planets as in Figure 5 after these bodies either grew to $1.0 M_\oplus$ and the primordial atmosphere was lost, or after 1 Gyr if the primordial atmosphere remained at the planet. Please note that this is no evolution but different outcomes, i.e., the small bullets do not evolve into the big ones (after Erkaev et al., 2023).

The magenta line in Figure 6, on the contrary, illustrates the heat production evolution of Earth. Although the specifics of the tectonic transitions are debated within the geodynamic community, it is common sense that different internal heating rates of terrestrial planets will result in different planetary evolutionary paths and tectonic regimes which can, for example, be observed by comparing Venus with Earth since it is suggested that Venus failed to develop accurate mantle convection and prolonged and continuous plate tectonics (Kaula, 1990; Kaula and Phillips, 1981; Weller et al., 2023) potentially due to the wrong abundance of heat-producing isotopes (*e.g.*, Kaula, 1999; O'Neill et al., 2014; Jellinek and Jackson, 2015; O'Neill et al., 2021).

The tectonic regime, therefore, determines the rate and style of cooling, the extent and mode of resurfacing, and strongly influences C-silicate and N-cycle, which are both needed for building up and maintaining secondary atmospheres, a temperate climate, and the longevity of stable environmental conditions relevant for the evolution of an Earth-like Habitat on which surficial complex aerobic life can evolve.

If the emergence of stable carbon-silicate and nitrogen cycles fails, one can expect that planets will not evolve $N_2$-$O_2$-dominated atmospheres with minor amounts of $CO_2$. In such a case, an Earth-like terrestrial planet that ends without a primordial atmosphere will most likely evolve a Venus- or Mars-like $CO_2$-dominated atmosphere (see Lammer et al. 2019 for a detailed discussion). The expected discovery of $CO_2$-dominated atmospheres with instruments such as JWST and ELT (*e.g.*, Snellen et al., 2015) and future space observatories like LIFE and HWO and, specifically, their derived frequency around rocky exoplanets in the HZCL will, therefore, allow for scientific assessment of $\beta_{env}^{cycles}$.

In Scherf *et al.* (2024, this issue), we apply Eqs. 2, 3, and 4 and will derive a maximum number of Earth-like Habitats within the galactic disk by only considering estimates for those constraints that are at least quantifiable to some certain extent within today's scientific knowledge and by ignoring those for which, at present, no quantitative estimates can be derived. We will, however, discuss the importance of further factors and how these might affect our derived maximum number, in Scherf *et al.* (2024, this issue) as well.

## 4. Conclusion

Since peculiar conditions are required for the origin and evolution of aerobic complex lifeforms, that is, millimeter-to-meter-sized advanced metazoans, we derived a formula that can be used to estimate the maximum number of potential Earth-like Habitats, $N_{EH}$, in the Galaxy, in which we substituted Eta-Earth, $\eta_\oplus$, with the occurrence rate of Earth-like Habitats, $\eta_{EH}$. The derived parameter $\eta_{EH}$ is more complex than the conventional $\eta_\oplus$, which only corresponds to the number of "terrestrial planets" in the HZ of solar-like stars. Our formulation minimizes, however, speculative assumptions and yields maximum results that are scientifically quantifiable and will be even more so in the near future. We also discussed in an exemplary manner two crucial, though often neglected, factors that are related to planetary accretion and the formation of primordial hydrogen-dominated atmospheres as well as to geophysical processes that are necessary for carbon-silicate and nitrogen-cycles to evolve and be stable over billions of years. These factors can be fine-tuned and/or constrained soon by the characterization of exoplanetary atmospheres with present and future space observatories and large ground-based telescopes such as JWST, LIFE, HWO, and the ELT. In Scherf *et al.* (2024, this issue), we will further estimate the maximum number of Earth-like Habitats within the galactic disk by applying our derived formula and by only considering relevant factors that are to some certain extent already scientifically quantifiable with our present knowledge. There, we will also discuss further, not yet quantifiable, requirements and assess implications that result from our results.

## Acknowledgments

The authors thank J. F. Kasting from the Department of Geosciences at Penn State University, USA, S. Mojzsis from the Research Centre for Astronomy and Earth Sciences (CSFK), Konkoly Observatory, Budapest, Hungary, and an anonymous referee for their suggestions and recommendations which helped to improve this work.

## Abbreviations

CAL: Complex Animal Life

CHEOPS: CHaracterising ExOPlanet Satellite

CNOPS: Carbon, Nitrogen, Oxygen, Phosphor, Sulphur

EH: Earth-like Habitat

ELT: Extremely Large Telescope

ETI: Extra-Terrestrial Intelligence

HWO: Habitable Worlds Observatory

HZ: Habitable Zone

HZCL: Habitable Zone for Complex Life

JWST: James Webb Space Telescope

LIFE: Large Interferometer for Exoplanets

SETI: Search for Extraterrestrial Intelligence

TESS: Transiting Exoplanet Survey Satellite

PLATO: PLAnetary Transits and Oscillation of stars

## Author Disclosure Statement

No competing financial interests exist.